\documentclass[useAMS,usenatbib,  usegraphicx, twocolumn]{mn2e}

\title[Stochastic oscillations of general relativistic disks]{Stochastic oscillations of general relativistic disks}
\author[T. Harko and G. Mocanu]{Tiberiu Harko$^1$ \thanks{E-mail:
harko@hkucc.hku.hk} and Gabriela Mocanu$^2$
\thanks{E-mail:gabriela.mocanu@ubbcluj.ro}\\
$^1$Department of Physics and Center for Theoretical and Computational Physics,\\
 University of Hong Kong,  Pok Fu Lam Road, Hong Kong, Hong Kong SAR, P. R. China\\
 $^2$Faculty of Physics, Department of Theoretical and Computational Physics, Babes-Bolyai University, Cluj-Napoca, Romania}

\begin{document}


\pagerange{\pageref{firstpage}--\pageref{lastpage}} \pubyear{2002}

\maketitle

\label{firstpage}

\begin{abstract}
We analyze the general relativistic oscillations of thin accretion disks around compact astrophysical objects interacting with the surrounding medium through non-gravitational forces. The interaction with the external medium (a thermal bath) is modeled via a friction force, and a random force, respectively. The general equations describing the stochastically perturbed disks are derived by considering the perturbations of trajectories of the test particles in equatorial orbits, assumed to move along the geodesic lines. By taking into account the presence of a viscous dissipation and of a stochastic force we show that the dynamics of the stochastically perturbed disks can be formulated in terms of a general relativistic Langevin equation. The stochastic energy transport equation is also obtained. The vertical oscillations of the disks in the Schwarzschild and Kerr geometries are considered in detail,    and they are analyzed by numerically integrating the corresponding Langevin equations. The vertical displacements, velocities and luminosities of the stochastically perturbed disks are explicitly obtained for both the Schwarzschild and the Kerr cases.
\end{abstract}


\section{Introduction}

The simplest physical model describing the random motion of a non-relativistic Brownian particle is given
by the Langevin equation, which can be written as \citep{Co04}
\begin{equation}
\frac{dp}{dt}=-\nu p+K+\xi (t),  \label{1}
\end{equation}
where $p$ is the (non-relativistic) momentum, and the three terms
of the right-hand side of Eq. (\ref{1}) correspond to the
friction, to an external force $K$, and to a stochastic force $\xi
$, respectively. $\nu $ is a coefficient called the friction
coefficient. Due to the friction term, the particle in Brownian
motion loses energy to the medium, but simultaneously gains energy
from the random kicks of the thermal bath, modeled by the random
force. The random
force, which we assume to be a white noise, satisfies the conditions $%
\left\langle \xi (t)\right\rangle =0$ and $\left\langle \xi (t)\xi \left(
t^{\prime }\right) \right\rangle =k\delta \left( t-t^{\prime }\right) $, respectively,
where $k$ is a constant. The separation of the force into frictional and
random parts is merely a phenomenological simplification - microscopically,
the two forces have the same origin (collision with the external medium constituents).
The Langevin equation has a large range of applications in physics, astrophysics,
engineering, biology etc. \citep{Co04}.

An exact formulation of the physics of Barnett relaxation, based on a realistic kinetic model of the relaxation mechanism which includes the alignment of the grain angular momentum in body coordinates by Barnett dissipation, disalignment by thermal fluctuations, and coupling of the angular momentum to the gas via gas damping was introduced in \citet{Laz}. The Fokker-Planck equation for the measure of internal alignment was solved by using numerical integration of the equivalent Langevin equation for Brownian rotation. It was also shown that in a steady state,
energy is transferred back to the grain material at an equal rate by Barnett dissipation.

A simple physical model to describe the dynamics of a massive point-like object, such as a black hole, near the center of a dense stellar system was developed in \citet{Chat}. The total force on the massive body can be separated into two independent parts, one of which is the slowly varying influence of the aggregate stellar system, and the other being the rapidly fluctuating stochastic force due to discrete encounters with individual stars. The motion of the black hole is  similar to that of a Brownian particle in a harmonic potential, and its dynamics can be analyzed by using an approach based on the  Langevin equation. By numerically solving the Langevin equation one can  obtain the average values, time-autocorrelation functions, and probability distributions of the black hole's position and velocity.  By using this model a lower limit on the mass of the black hole Sgr A* in the Galactic center can be derived.

A Langevin--type treatment of the motion of a charged particle in the inter-galactic medium magnetic field, which allows to estimate both the average and of the root mean square time-delay for particles of given energy, was introduced in \citet{Vietri}. This model was compared  with a scenario where the particles are accelerated at internal shocks.
The generalized thermodynamics and the collapse of a system of self-gravitating Langevin particles exhibiting anomalous diffusion in a space of dimension D was considered in \citet{Cha}. The equilibrium states correspond to polytropic distributions. The dynamical stability of stellar systems, gaseous stars and two-dimensional
vortices was studied by using a thermodynamical analogy.

There are many processes in the accretion of material onto
an active galactic nuclei that occur on relatively short timescales as compared
to the cooling time of the X-ray emitting gas in and around
elliptical galaxies and galaxy clusters.  Accretion rates onto active galactic nuclei are likely to be extremely variable on short timescales. Using a Langevin type equation \citet{Pope} has shown that, for a simple feedback
system, this can induce variability in the active galactic nuclei power output that is of much larger amplitude,
and persists for longer timescales, than the initial fluctuations. An implication
of this result is that rich galaxy clusters are expected to show the largest and longest-lived
fluctuations. Stochastic variations in the accretion rate also mean that the active galactic nuclei injects
energy across a wide range of timescales.

Accretion disks around compact objects have become standard models for a
number of astrophysical phenomena like X-ray binaries or active galactic
nuclei. The disks are usually considered as being composed from massive test particles that move
in the gravitational field of the central compact object.  Waves and normal-mode oscillations in geometrically thin and thick disks around compact objects have
been studied extensively both within Newtonian gravity (see \citet{Kato1}
2001 for a review) and within a relativistic framework \citep{Okaz, Per, Sem, Kato1, Sil, Rod,Rez, Zan, Bla, On, Shi1}. The oscillations of the accretion disks  may produce the quasi-periodic oscillations (QPO's) \citep{Tit,Tag, Fan, Hor} in black hole low-mass X-ray binaries.

It is the purpose of the present paper to propose a fully
general relativistic model for the description of the oscillations of the particles in the  accretion disks
around compact objects in contact with an external heat bath.  The oscillations of the disk, assumed to interact
with an external medium (the heat bath), are described by a general
relativistic Langevin equation, which is derived, for the case of a general axisymmetric gravitational
field, by considering the perturbations of the geodesic equation describing the motion of test particles in stable circular orbits. The energy transport in the presence of the stochastic fluctuations of the disk is discussed in detail. As an application of our general formalism we consider in detail the vertical stochastic oscillations of the accretion disks in both the static Schwarzschild and rotating Kerr geometries. In these cases the equations of motion of the disk can be formulated as Langevin equations, describing the interaction between the disk and an external thermal bath. By numerically integrating the corresponding Langevin equation we obtain the vertical displacements, velocities, and the luminosities of the stochastically oscillating disks.

The present paper is organized as follows. In Section~\ref{s2} we briefly review the motion of the test particles in axisymmetric gravitational fields. In Section~\ref{s3} we consider the stochastic perturbations of the geodesic trajectories of test particles moving around a central compact object.  The general system of perturbed equations for stochastic disks is obtained in Section~\ref{s4}. The energy transfer in stochastically perturbed disks is discussed in Section~\ref{s5}. The vertical motions of the disks are considered in Section~\ref{s6}, and it is shown that they can be described by a Langevin type equation. The disk oscillation frequencies are also obtained for both the Schwarzschild and the Kerr metrics. The vertical oscillations of the stochastically perturbed disks are considered in Section~\ref{s7}, and the vertical displacements, velocities and luminosities are obtained, by numerically integrating the corresponding Langevin equation, for both the static Schwarzschild and rotating Kerr geometries. We discuss and conclude our results in Section~\ref{s8}.

\section{ Motion of test particles in stationary axisymmetric gravitational
fields}\label{s2}

The metric generated by a rotating axisymmetric compact general relativistic
object can be generally given in cylindrical coordinates $\left( ct,\rho
,\phi ,z\right) $ as
\begin{equation}
ds^{2}=g_{tt}\left( cdt\right) ^{2}+2g_{t\phi }cdtd\phi +g_{\phi \phi }d\phi
^{2}+g_{\rho \rho }\left( d\rho ^{2}+dz^{2}\right) ,
\end{equation}
where all the metric functions are functions of $\rho $ and $z$ only. The
equatorial plane is placed at $z=0$. For a given timelike worldline $x^{\mu
}\left( s\right) $ with four velocity $u^{\mu }=dx^{\mu }/ds$ and azimuthal
angular velocity $\Omega =d\phi /dt$, the specific azimuthal angular
momentum $\tilde{L}$ of a particle of mass $m$, given by
\begin{equation}
\tilde{L}=u_{\phi }=u^{t}\left( g_{t\phi }+g_{\phi \phi }\Omega /c\right) ,
\label{3}
\end{equation}
and the specific energy $\tilde{E}$, given by
\begin{equation}
\tilde{E}=-u_{t}=-u^{t}\left( g_{tt}+g_{t\phi }\Omega /c\right) ,  \label{4}
\end{equation}
are constants of motion. The four-velocity satisfies the condition $u_{\mu
}u^{\mu }=-1$. In a stationary axisymmetric gravitational field the most
important types of worldlines correspond to spatially circular orbits, given
by $\rho =$ constant, $z=$ constant and $\Omega =$ constant, respectively.
In this case the four-velocity is given by
\begin{equation}
u^{\mu }=u^{t}\left( 1,0,\Omega /c,0\right) ,
\end{equation}
where
\begin{equation}\label{utt}
u^{t}=\frac{1}{\sqrt{-g_{tt}-2g_{t\phi }\left(\Omega /c\right)-g_{\phi \phi }\left(\Omega /c\right)^{2}}}%
=\left( \tilde{E}-\frac{\Omega }{c}\tilde{L}\right) ^{-1}.
\end{equation}

For circular orbits the four-acceleration $a_{\mu }=-(1/2)g_{\alpha \beta ,\mu} u^{\alpha }u^{\beta }$ of the test particles has only a radial component,
\begin{eqnarray}
a_{\rho }&=&-\frac{1}{2}g_{\alpha \beta ,\rho} u^{\alpha }u^{ \beta }=\nonumber\\
&&-\left(\frac{1}{2}g_{tt,\rho }+g_{t\phi ,\rho}\frac{\Omega}{c}+\frac{1}{2}g_{\phi \phi,\rho}\frac{\Omega ^2}{c^2}\right)\left(u^t\right)^2.
\end{eqnarray}

 The radial component of the force $a_{\rho }$ vanishes for two particular values $\Omega _{\pm }$ of the angular velocity, given by
\begin{equation}
\frac{\Omega _{\pm }}{c}=\frac{-g_{t\phi ,\rho }\pm \sqrt{g_{t\phi ,\rho
}^{2}-g_{tt,\rho }g_{\phi \phi ,\rho }}}{g_{\phi \phi ,\rho }}.  \label{om}
\end{equation}

\section{Stochastic perturbation of an equatorial orbit}\label{s3}

In the absence of any external perturbation the particles in the disk move
along the geodesic lines given by
\begin{equation}
\frac{d^{2}x^{\mu }}{ds^{2}}+\Gamma _{\alpha \beta }^{\mu }\frac{dx^{\alpha }%
}{ds}\frac{dx^{\beta }}{ds}=0,
\end{equation}
where $\Gamma _{\alpha \beta }^{\mu }$ are the Christoffel symbols
associated to the metric. If the disk is perturbed, a point particle in the
nearby position $x^{\prime \mu }$  must also satisfy a generalized
geodesic equation
\begin{equation}
\frac{d^{2}x^{\prime \mu }}{ds^{2}}+\Gamma _{\alpha \beta }^{\prime \mu }%
\frac{dx^{\prime \alpha }}{ds}\frac{dx^{\prime \beta }}{ds}=\frac{1}{m}%
f^{\mu },
\end{equation}
where $f^{\mu }$ is the external, non-gravitational force, acting on the
particle. The coordinates $x^{\prime \mu }$ are given by $x^{\prime \mu
}=x^{\mu }+\delta x^{\mu }$, where $\delta x^{\mu }$ is a small quantity.
Thus, in the first approximation, we obtain for the Christoffel symbols
\begin{equation}
\Gamma _{\alpha \beta }^{\prime \mu }\left( x+\delta x\right) =\Gamma
_{\alpha \beta }^{\mu }\left( x\right) +\Gamma _{\alpha \beta ,\lambda
}^{\mu }\delta x^{\lambda },
\end{equation}
where $\Gamma _{\alpha \beta ,\lambda }^{\mu }=\partial \Gamma _{\alpha
\beta }^{\mu }/\partial x^{\lambda }$. Therefore the equation of motion for $\delta
x^{\mu }$ becomes
\begin{equation}
\frac{d^{2}\delta x^{\mu }}{ds^{2}}+2\Gamma _{\alpha \beta }^{\mu }\frac{%
dx^{\alpha }}{ds}\frac{d\delta x^{\beta }}{ds}+\Gamma _{\alpha \beta ,\lambda
}^{\mu }\frac{dx^{\alpha }}{ds}\frac{dx^{\beta }}{ds}\delta x^{\lambda
}=f^{\mu }.
\end{equation}

If $f^{\mu }=0$ we reobtain the equation of perturbation of the geodesic line introduced in \citet{Shi}. In the following we introduce the four-velocity of the perturbed motion as $%
\delta V^{\mu }=d\delta x^{\mu }/ds$. $\delta V^{\mu }$ satisfies the condition $u_{\mu
}\delta V^{\mu }=0$. We consider that the particles in the disk are in contact with
an isotropic and homogeneous heat bath. The interaction of the particles
with the bath is described by a friction force, and a random force. With
respect to a comoving observer frame the heat bath has a non-vanishing
average four-velocity $U^{\mu }$.

In the non-relativistic case the friction force is given by $%
f_{fr}^{i}=-m\nu v^{i}$, where $\nu $ is the friction coefficient, and $%
v^{i} $ are the components of the non-relativistic velocity. The
relativistic generalization of the friction force requires the introduction
of the friction tensor $\nu _{\alpha }^{\mu }$, given by \citep{Du05a, Du05b, Du09}
\begin{equation}
\nu _{\alpha }^{\mu }=\nu m\left( \delta _{\alpha }^{\mu }+\delta V_{\alpha }\delta V^{\mu
}\right) .
\end{equation}

The friction force can be expressed then as
\begin{eqnarray}
f_{fr}^{\mu }&=&-\nu _{\alpha }^{\mu }\left( \delta V^{\alpha }-U^{\alpha }\right)
=\nonumber\\
&&-\nu m\left( \delta _{\alpha }^{\mu }+\delta V_{\alpha }\delta V^{\mu }\right) \left(
\delta V^{\alpha }-U^{\alpha }\right) .
\end{eqnarray}

Let us now introduce a Gaussian stochastic vector field $m\xi ^{\mu }\left[
g;x\right] $, where $g$ is the determinant of the metric tensor, defined by
the following correlators: $\left\langle \xi ^{\mu }\left[ g;x\right]
\right\rangle =0$ and $\left\langle \xi ^{\mu }\left[ g;x\right] \xi ^{\nu }%
\left[ g;y\right] \right\rangle =D^{\mu \nu }\left[ g;x;y\right] $, where $%
D^{\mu \nu }\left[ g;x;y\right] $ is the noise kernel tensor. From a physical point of view $m\xi ^{\mu }\left[
g;x\right] $ represents the stochastic force generated by the interaction of the particles of the disk with the external heat bath. Thus the
motion of the general relativistic massive test particles in the stochastically perturbed disk can be
described by the following general relativistic Langevin type equation
\begin{eqnarray}
&&\frac{d\delta V^{\mu }}{ds}+2\Gamma _{\alpha \beta }^{\mu }u^{\alpha }\delta V^{\beta
}+\Gamma _{\alpha \beta ,\lambda }^{\mu }u^{\alpha }u^{\beta }\delta
x^{\lambda }=\nonumber\\
&&-\nu \left( \delta _{\alpha }^{\mu }+\delta V_{\alpha }\delta V^{\mu }\right)
\left( \delta V^{\alpha }-U^{\alpha }\right) +\xi ^{\mu }\left[ g;x\right] .
\label{lange}
\end{eqnarray}

In the following we will use Eq.~(\ref{lange}) to analyze the physical properties of the randomly fluctuating accretion disks.

\section{Stochastic oscillations of accretion disks}\label{s4}

We assume that the heat bath is at rest, so that $U^{\mu }=\left(
g_{tt}^{-1/2},0,0,0\right) $. Also we assume that, since the perturbations
are small, the variation of the gravitational field along the $z$ -
direction can be neglected. This implies that along the equatorial plane $%
g_{\mu \nu ,z}=0$, but $g_{\mu \nu ,zz}\neq 0$. Then from the Langevin
equation Eq.~(\ref{lang}) we obtain the equations of motion of the particles in the stochastically perturbed disk as
\begin{eqnarray}
&&\frac{d^{2}\delta t}{ds^{2}}+2\left( \Gamma _{t\rho }^{t}+\Gamma _{\phi \rho
}^{t}\frac{\Omega }{c}\right) u^{t}\frac{d\delta \rho }{ds}=\nonumber\\
&&-\nu \left( \delta
_{\alpha }^{t}+\delta V_{\alpha }\delta V^{t}\right) \left( \delta V^{\alpha }-U^{\alpha }\right)
+\xi ^{t}\left[ g;x\right] ,
\end{eqnarray}
\begin{eqnarray}
&&\frac{d^{2}\delta \rho }{ds^{2}}+2\left( \Gamma _{tt}^{\rho }+\Gamma
_{t\phi }^{\rho }\frac{\Omega }{c}\right) u^{t}\frac{d\delta \rho }{ds}+ \nonumber\\
&&2\left( \Gamma
_{t\phi }^{\rho }+\Gamma _{\phi \phi }^{\rho }\frac{\Omega }{c}\right) u^{t}\frac{%
d\delta \phi }{ds}+ \nonumber\\
&&\left[ \Gamma _{tt,\rho }^{\rho }+2\Gamma _{t\phi ,\rho }^{\rho }\frac{\Omega }{c}
+\Gamma _{\phi \phi ,\rho }^{\rho }\left(\frac{\Omega }{c}\right)^{2}\right] \left( u^{t}\right)
^{2}\delta \rho =\nonumber\\
&&-\nu \left( \delta _{\alpha }^{\rho}+\delta V_{\alpha }\delta V^{\rho}\right)
\delta V^{\alpha }+\xi ^{\rho}\left[ g;x\right] ,
\end{eqnarray}
\begin{eqnarray}
&&\frac{d^{2}\delta \phi }{ds^{2}}+2\left( \Gamma _{t\rho }^{\phi }+\Gamma
_{\phi \rho }^{\phi }\frac{\Omega }{c}\right) u^{t}\frac{d\delta \rho }{ds}=\nonumber\\
&&-\nu
\left( \delta _{\alpha }^{\phi }+\delta V_{\alpha }\delta V^{\phi }\right) \delta V^{\alpha }+\xi
^{\phi }\left[ g;x\right] ,
\end{eqnarray}
\begin{eqnarray}
&&\frac{d^{2}\delta z}{ds^{2}}+\left[ \Gamma _{tt,z}^{z}+2\Gamma _{t\phi
,z}^{z}\frac{\Omega }{c}+\Gamma _{\phi \phi ,z}^{z}\left(\frac{\Omega }{c}\right)^{2}\right] \left(
u^{t}\right) ^{2}\delta z=\nonumber\\
&&-\nu \left( \delta _{\alpha }^{z}+\delta V_{\alpha
}\delta V^{z}\right) \delta V^{\alpha }+\xi ^{z}\left[ g;x\right] .
\end{eqnarray}

In the absence of the friction force and of the stochastic force generated by the interaction between the disk and the heat bath we reobtain the equations for the free oscillations of accretion disks \citep{Sem}.

\section{Energy transfer in stochastically oscillating disks}\label{s5}

In the following we will use the Eckart model for dissipative processes \citep{Wein}, and we will choose a frame defined by the family of observers moving with
the normalized four-velocity $ v^{\mu }=n^{\mu }/n$ parallel to the oscillating matter fluid of the disk, where $n^2=-g_{\mu \nu}n^{\mu }n^{\nu }$.

We assume that the stochastically oscillating disk consists of at
least one fluid, whose particle number density flux $n^{\mu }=n v^{\mu }$ is conserved, $n^{\mu }_{;\mu }=0$, where ; denotes the covariant derivative with respect to the metric, and $n$ is the particle number density. Let us now introduce a Gaussian stochastic tensor field $\xi _{|mu \nu}[g; x]$ defined by the  correlators $\langle \xi _{\mu \nu}[g; x]\rangle =0$ and $\langle \xi _{\alpha \beta}[g; x]\xi _{|mu \nu}[g; y]\rangle=N_{\alpha \beta \mu \nu }[g;x;y]$, respectively, where $\langle\rangle$ means statistical average \citep{stoch}. The symmetry and positive semi-definite property of the noise kernel guarantees that the stochastic field tensor $\xi _{\mu \nu }[g;x]$, or $\xi _{\mu \nu }$ for short, is well-defined. We assume that this general relativistic stochastic tensor describes  the fluctuations of the energy-momentum tensor $\delta T^{\mu \nu }$ of the matter in the disk due to the interaction with the thermal bath. Hence  $\delta T^{\mu \nu }$  can be written as
\begin{equation}
\delta T^{\mu \nu }=\rho  v^{\mu } v^{\nu }+2q^{(\mu } v^{\nu )}+\pi ^{\mu \nu }+\xi ^{\mu \nu},
\end{equation}
where $\rho $ is the energy density, $q^{\mu }$ is the transverse momentum, and $\pi ^{\mu \nu }$ is the anisotropic stress tensor, as measured by an observer moving with the particle flux. $q^{\mu }$ and $\pi ^{\mu \nu }$ satisfy the relations $q^{\mu }v_{\mu }=0$ and $\pi ^{\mu \nu }v_{\mu }=0$, respectively. Within the framework of this physical interpretation we can define the internal energy $\epsilon $ through the relation $n(\epsilon + \epsilon _0) = \rho$, where $\epsilon _0$ is an arbitrary constant \citep{Hawk}. The energy-momentum tensor is covariantly conserved, so that $\delta T^{\mu \nu }_{;\mu }=0$.
By contracting $\delta T^{\mu \nu }$ with the observer's four velocity $v_{\mu}$  we obtain
\begin{equation}\label{eccons}
\left(v_{\mu }\delta T^{\mu \nu }\right)_{;\nu }=v_{\mu ;\nu}\delta T^{\mu \nu}.
\end{equation}
For the first term in Eq.~(\ref{eccons}) we obtain
\begin{equation}
 v_{\mu }\delta T^{\mu \nu }=-\rho v^{\nu}-q^{\nu}+v_{\mu }\xi ^{\mu \nu}.
 \end{equation}

 By using the definition of the internal energy and the conservation of the particle number density
flux we obtain
 \begin{equation}
 \left(v_{\mu }\delta T^{\mu \nu }\right)_{;\nu }=-n\dot{\epsilon}-q^{\mu}_{;\mu} +v_{\mu ;\nu}\xi ^{\mu \nu}+v_{\mu }\xi ^{\mu \nu}_{;\nu },
 \end{equation}
 where we have denoted $\dot{\epsilon}=\epsilon _{;\nu }v^{\nu}$. The right-hand side in Eq.~(\ref{eccons}) can be written as
 \begin{equation}
 v_{\mu ;\nu}\delta T^{\mu \nu}=q^{\mu }\dot{v}_{\mu }+v_{\mu ;\nu}\pi ^{\mu \nu} +v_{\mu ;\nu}\xi ^{\mu \nu}.
 \end{equation}

 Thus, the local form of the energy balance  of the stochastically oscillating disk is given, from the fluid observer's point of view, by
\begin{equation}
 n\dot{\epsilon}+q^{\mu}_{;\mu}+q^{\mu }\dot{v}_{\mu }+v_{\mu ;\nu}\pi ^{\mu \nu}=v_{\mu }\xi ^{\mu \nu}_{;\nu }.
 \end{equation}

 When the stochastic term in the energy-momentum tensor is ignored, and by assuming that in this case $\delta T_{\mu \nu}$ takes the form $\delta T_{\mu \nu}=\left(\rho +p\right)v^{\mu }v^{\nu}+pg^{\mu \nu}$, we can define the hydrostatic pressure $p$ of the fluid as the trace of the stress tensor $\pi ^{\mu \nu}$, $p=(1/3)\pi ^{\mu }_{\mu }$. This allows us to define the viscous stress tensor $P^{\mu \nu}$ as \citep{Wein}
 \begin{equation}
 P^{\mu \nu}=-\pi ^{\mu \nu}+ph^{\mu \nu},
 \end{equation}
 where $h^{\mu \nu}$ is the orthogonal projection operator to the observer's four-velocity defined as $h^{\mu \nu}=g^{\mu \nu}+v^{\mu }v^{\nu }$. The viscous stress tensor can be obtained in a general form as
 \begin{equation}
 P^{\mu \nu}=\lambda \left[\sigma ^{\mu \nu}+2\dot{v}^{(\mu}v^{\nu )}\right],
 \end{equation}
 where $\lambda $ is the viscosity coefficient and $\sigma _{\mu \nu}=v_{(\mu ;\nu )}-(1/3) v^{\alpha }_{;\alpha }h_{\mu \nu }$. By defining the invariant specific volume $v$ as the inverse of the particle number density $n$ it follows that $h^{\mu \nu}v_{\mu ;\nu}=v_{\mu }^{\mu}=n\dot{v}$. Then for the equation of the local energy balance we obtain
 \begin{equation}\label{enfin}
 n\left(\dot{\epsilon}+p\dot{v}\right)+q^{\mu }_{;\mu}+q^{\mu }\dot{v}_{\mu }-v_{\mu ;\nu}P^{\mu \nu }=v_{\mu }\xi ^{\mu \nu}_{;\nu }.
 \end{equation}

As for the transverse momentum $q^{\mu }$ it can be obtained from the relativistic analogue of Fourier's law \citep{Wein}, namely
\begin{equation}
q^{\mu }=-\kappa h^{\mu \nu}\left(T_{;\nu}+T\dot{v}_{\nu }\right),
\end{equation}
where $\kappa $ is the thermal conductivity coefficient and $T$ is the equilibrium temperature of the system. Eq.~(\ref{enfin}) is analogous to the non-relativistic energy balance equation in the presence of a stochastic force $\vec{\xi }$, given by
\begin{equation}\label{nrel}
n\frac{d\epsilon}{dt}+\nabla \cdot\vec{q}-\left(\vec{\Pi }\cdot\nabla\right)\cdot\vec{v}=\vec{v}\cdot\vec{\xi},
\end{equation}
where $\epsilon $ is the internal energy, $\vec{q}$ is the heat flux, $\vec{\Pi }$ is the total stress tensor, and $\vec{v}$ is the fluid's three velocity. In Eq.~(\ref{enfin})  only the term containing the four-acceleration $\dot{u}_{\mu }$  does not have a Newtonian counterpart. When $\vec{\xi}\equiv 0$, Eq.~(\ref{nrel}) reduces to the standard energy balance equation in non-relativistic fluid mechanics \citep{Lan}.

\section{Equation of motion of vertically perturbed accretion disks}\label{s6}

In the following we will consider only the vertical oscillations of the
disk, which, for an arbitrary axisymmetric metric, are described by the equation
\begin{eqnarray}
&&\frac{d^{2}\delta z}{ds^{2}}+\left[ \Gamma _{tt,z}^{z}+2\Gamma _{t\phi
,z}^{z}\frac{\Omega }{c}+\Gamma _{\phi \phi ,z}^{z}\left(\frac{\Omega }{c}\right)^{2}\right] \left(
u^{t}\right) ^{2}\delta z=\nonumber\\
&&-\nu \left( \delta _{\alpha }^{z}+\delta V_{\alpha
}\delta V^{z}\right) \delta V^{\alpha }+\xi ^{z}\left[ g;z\right] .
\end{eqnarray}

Since $\delta V^{\mu }$ is a small quantity, the equation of motion can be further
simplified to
\begin{equation}\label{lang}
\frac{d^{2}\delta z}{ds^{2}}+\nu \frac{d\delta z}{ds}+\omega _{\perp
}^{2}\delta z=\xi ^{z}\left[ g;z\right] ,
\end{equation}
where
\begin{equation}
\omega _{\perp }^{2}=\left[ \Gamma _{tt,z}^{z}+2\Gamma _{t\phi
,z}^{z}\frac{\Omega }{c}+\Gamma _{\phi \phi
,z}^{z}\left(\frac{\Omega }{c}\right)^{2}\right] \left( u^{t}\right) ^{2}.
\end{equation}
The angular frequency
with respect to the radial infinity is obtained as
\begin{equation}
\Omega _{\perp
}^{2}=\omega _{\perp }^{2}/\left( u^{t}\right) ^{2}=\Gamma
_{tt,z}^{z} +2\Gamma _{t\phi
,z}^{z}\frac{\Omega }{c}+\Gamma _{\phi \phi ,z}^{z}\left(\frac{\Omega }{c}\right)
^{2}.
\end{equation}

In the following we consider the vertical oscillations of the disks around black holes described by the Schwarzschild and the Kerr geometry, respectively.

\subsection{Disk oscillation frequencies in the Schwarzschild geometry}

The general form of a static axisymmetric metric can be given as \citep{Bi05}
\begin{equation}
ds^{2}=-e^{2U}c^{2}dt^{2}+e^{-2U}\left[ e^{2\gamma }\left( d\rho
^{2}+dz^{2}\right) +\rho ^{2}d\phi ^{2}\right] .
\end{equation}
For the Schwarzschild solution
\begin{equation}
U=\frac{1}{2}\ln \frac{\sqrt{\rho ^{2}+\left( M+z\right) ^{2}}+\sqrt{\rho ^{2}+\left(
M-z\right) ^{2}}-2M}{\sqrt{\rho ^{2}+\left( M+z\right) ^{2}}+\sqrt{%
\rho ^{2}+\left( M-z\right) ^{2}}+2M},
\end{equation}
and
\begin{equation}
\gamma =\frac{1}{2}\ln \frac{\left[ \sqrt{\left( M+z\right) ^{2}+\rho ^{2}}+%
\sqrt{\left( M-z\right) ^{2}+\rho ^{2}}\right] ^{2}-4M^{2}}{4\sqrt{\left(
M+z\right) ^{2}+\rho ^{2}}\sqrt{\left( M-z\right) ^{2}+\rho ^{2}}},
\end{equation}
respectively, where $M$ is the mass of the central compact object in geometrical units \citep{Bi05}.
The usual Schwarzschild line element in $\left( ct,r,\theta ,\phi \right) $
coordinates is recovered by making the transformations $\rho =\sqrt{r^{2}-2Mr%
}\sin \theta $ and $z=\left( r-M\right) \cos \theta $.


For an accretion disk in the Schwarzschild geometry the radial frequency at infinity and the proper frequency are given by
\begin{equation}
\Omega _{\perp }^{2}\left(\rho,z,M\right)=\frac{e^{4U-2\gamma }}{1-\rho U_{,\rho }}U_{,zz},
\end{equation}
and
\begin{equation}
\omega _{\perp }^{2}\left(\rho,z,M\right)=\frac{e^{2U-2\gamma }}{1-2\rho U_{,\rho }}U_{,zz},
\label{freq}
\end{equation}
respectively.
Estimating $\Omega _{\perp }^{2}$ in the equatorial plane at $z=0$ gives $\Omega _{\perp
}^{2}|_{z=0}=M/\left( M+\sqrt{M^{2}+\rho ^{2}}\right)^3 $, while for $\omega _{\perp}^2$ we obtain
\begin{equation}\label{Schfr}
\omega _{\perp }^{2}\left(M,\rho \right)=\frac{M}{\rho ^2\sqrt{M^2+\rho ^2}-2M^2\left(M+\sqrt{M^2+\rho ^2}\right)}.
\end{equation}


\subsection{Disk oscillation frequencies in the Kerr geometry}

The line element of a stationary axisymmetric spacetime is given by the
Lewis-Papapetrou metric \citep{Bi05}
\begin{equation}
ds^{2}=e^{2\left( \gamma -\psi \right) }\left( d\rho ^{2}+dz^{2}\right)
+\rho ^{2}e^{-2\psi }d\phi ^{2}-e^{2\psi }\left( cdt-\omega d\phi \right)
^{2},
\end{equation}
where $\gamma $ and $\psi $ are functions of $z$ and $\rho $ only. The
metric functions generating the Kerr black hole solution in axisymmetric
form are given by
\begin{equation}
\psi =\frac{1}{2}\ln \frac{\left( R_{+}+R_{-}\right) ^{2}-4M^{2}+\left(
a^{2}/\sigma ^{2}\right) \left( R_{+}-R_{-}\right) ^{2}}{\left(
R_{+}+R_{-}+2M\right) ^{2}+\left( a^{2}/\sigma ^{2}\right) \left(
R_{+}-R_{-}\right) ^{2}},
\end{equation}
\begin{equation}
\gamma =\frac{1}{2}\ln \frac{\left( R_{+}+R_{-}\right) ^{2}-4M^{2}+\left(
a^{2}/\sigma ^{2}\right) \left( R_{+}-R_{-}\right) ^{2}}{4R_{+}R_{-}},
\end{equation}
\begin{equation}
\omega =-\frac{aM}{\sigma ^{2}}\frac{\left( R_{+}+R_{-}+2M\right) \left[
\left( R_{+}-R_{-}\right) ^{2}-4\sigma ^{2}\right] }{\left(
R_{+}+R_{-}\right) ^{2}-4M^{2}+\left( a^{2}/\sigma ^{2}\right) \left(
R_{+}-R_{-}\right) ^{2}},
\end{equation}
where $M$ and $a$ are the mass and the specific angular momentum of the
compact object, $R_{\pm }=\sqrt{\rho ^{2}+\left( z\pm \sigma \right) ^{2}}$%
and $\sigma =\sqrt{M^{2}-a^{2}}$. The usual form of the Kerr line element in
Boyer-Lindquist coordinates $\left( t,r,\theta ,\phi \right) $ is recovered
by performing the coordinates transformation $\rho =\sqrt{r^{2}-2Mr+a^{2}}%
\sin \theta $ and $z=\left( r-M\right) \cos \theta $, respectively.

The frequency of the disk oscillations in the Kerr geometry
is given by
\begin{equation}\label{Kerrfr}
\omega _\perp ^2 \left(\rho , M,a\right)= \frac{M \left(\varpi _1a+\varpi _3\right)}{\varpi_2a+\varpi _4},
\end{equation}
where we have denoted
\begin{eqnarray}
\varpi _1&=&4 M^{3/2} R_{\sigma} \delta_-^3 \delta_+^{13/2}\times \nonumber\\
&& \left[8 M (2 M + R_{\sigma})
\delta_+ +
   2 (5 M + 3 R_{\sigma}) \rho^2\right],
\end{eqnarray}
\begin{eqnarray}
\varpi _2=-8 M^{3/2} R_{\sigma}  \delta_-^3 \delta_+^{19/2} (4 M \delta_+ +
   2 R_{\sigma} \delta_+ + \rho^2),
\end{eqnarray}
\begin{eqnarray}
\varpi_3&=& 4 R_{\sigma} \delta_-^3 \delta_+^6 \left[2 (8 M^3 - 3 M R_{\sigma}^2 -
      R_{\sigma}^3) \delta_+^2 \right]+\nonumber \\
      && 4 R_{\sigma} \delta_-^3 \delta_+^6 \left[ 3 (6 M^2 + 3 M R_{\sigma} + R_{\sigma}^2) \delta_+ \rho^2 +
   3 M \rho^4\right],\nonumber\\
\end{eqnarray}
and
\begin{eqnarray}
\varpi _4 &=& -4R_{\sigma} \delta_-^3 \delta_+^{10}\times \nonumber\\
 &&\left[(8 M^3 - 3 M R_{\sigma}^2 - R_{\sigma}^3)
\delta_+ +
   3 M (2 M + R_{\sigma}) \rho^2\right],\nonumber\\
\end{eqnarray}
respectively, with $\delta _{\pm}=\sqrt{\rho ^2+\sigma ^2}\pm M$ and $R_{\sigma }=\sqrt{\rho ^2+\sigma ^2}$, respectively.
In the limit $a=0$, $R_{\sigma }=\sqrt{M^2+\rho ^2}$ and $\delta _{\pm }=R_{\sigma }\pm M$,  and after some algebraic manipulation,
we recover Eq.~(\ref{Schfr}) giving the oscillation frequency of the Schwarzschild disk.

\section{Vertical stochastic oscillations of perturbed accretion disks}\label{s7}

By assuming that the velocity of the perturbed disk oscillations is small,
we can approximate $ds$ as $ds\approx cdt$, where $t$ is the time
coordinate. We also assume that the external perturbative force is a
function of time only, and neglect its possible spatial coordinate
dependence. In this case the equation of motion of the disk in contact with
a heat bath follows from Eq.~(\ref{lang}) and  is given by
\begin{equation}\label{eqint1}
\frac{d^{2}\delta z}{dt^{2}}+c\nu \frac{d\delta z}{dt}+c^{2}\omega _{\perp
}^{2}\left(\rho, M,a\right)\delta z=c^{2}\xi ^{z}\left( t\right) ,
\end{equation}%
where the oscillation frequency of the disk is given by Eq.~(\ref{Schfr}) in the
Schwarzschild case, and by Eq.~(\ref{Kerrfr}) in Kerr case, respectively. The random force $\xi ^{z}\left( t\right)$ has a zero mean and a variance $\langle \xi ^{z}\left( t\right)\xi ^{z}\left( t'\right)\rangle=\left(D/c^2\right)\delta \left(t-t'\right)$, where $\delta \left(t-t'\right)$ is the Dirac delta function and $D\geq 0$.   By introducing the velocity $\delta V^z=d\delta z/\delta t$ of the vertical disk oscillation, Eq.~(\ref{eqint1}) can be written in the form
\begin{equation}\label{eqint2}
d\delta z=\delta V^zdt,
\end{equation}
\begin{equation}\label{eqint2a}
d\delta V^z=-c\nu \delta V^zdt-c^2\omega _{\perp
}^{2}\left(\rho, M\right)\delta z dt+D^{1/2}dW(t),
\end{equation}
where $W(t)$ is a collection of standard Wiener processes.

A Wiener process $W(t)\geq 0$, is a one-parameter
family of Gaussian random variables, with expectations zero and covariances $E\left(W(s)W(t)\right)={\rm min} \{s,t\}$. Because the $W(t)$ are all Gaussian, this information suffices to determine joint probabilities. Alternatively, $W(t)$ may be viewed as a 'random'continuous function with $W(0)=0$. A Wiener process may be
generated at consecutive grid points $t_n$ by $W(0)=0$, $W\left(t_n\right)=W\left(t_{n-1}\right)+\left(t_n-t_{n-1}\right)Z_n$, and $\{Z_n\}$ is a set of independent standard Gaussian random variables (with mean 0 and variance
1) \citep{Skeel}.

In order to numerically integrate Eqs.~(\ref{eqint2}) and (\ref{eqint2a}) we consider the  Br\"unger - Brooks - Karplus (BBK) scheme \citep{bbk}. If the time step is taken as $\Delta t$, we denote $\delta z_{n}\approx \delta z\left(n\Delta t\right)$, and we define $f_n=-c^2\omega _{\perp
}^{2}\left(\rho, M,a\right)\delta z_n \Delta t^2$. The recurrence relation for the BBK integrator is
\begin{eqnarray}
\delta z_{n+1}&=&\frac{2}{1+g/2}\delta z_n-\frac{1-g/2}{1+g/2}\delta z_{n-1}+\frac{1}{1+g/2}f_n+\nonumber\\
&&\frac{1}{1+g/2}\left(D\Delta t^3\right)^{1/2}Z_n,
\end{eqnarray}
where we have denoted $g=c\nu \Delta t$.  The starting procedure of the BBK integrator is
given by
\begin{equation}
\delta z_1=\delta z_0+\left(1-\frac{g}{2}\right)V_0\Delta t+\frac{1}{2}f_0+\left(\frac{D\Delta t ^3}{4}\right)^{1/2}Z_0,
\end{equation}
where $\delta z_0$ and $V_0$ are the initial displacement and velocity, respectively.
The velocity $\delta V^z$ can be obtained as
\begin{equation}
\delta V^{z}_{n}=\frac{\delta z_n-\delta z_{n-1}}{\Delta t}.
\end{equation}

By defining
\begin{equation}
E=\frac{1}{2}\left( \frac{d\delta z}{dt}\right) ^{2}+\frac{1}{2}c^{2}\omega _{\perp
}^{2}\left(\rho, M,a\right)\delta z^2,
\end{equation}%
as the total energy per unit mass of the oscillating disk, the luminosity per unit mass $L$ of the disk,
representing the energy lost by the disk due to viscous dissipation and to
the presence of the random force, is given by
\begin{equation}
L=-\frac{dE}{dt}=c\nu \left( \frac{d\delta z}{dt}\right) ^{2}-c^2\frac{d\delta z}{dt}\xi ^{z}\left( t\right) .
\end{equation}

\subsection{The Schwarzschild stochastic disk}

In the case of an accretion disk in the static Schwarzschild geometry of a black hole, the proper frequency of the disk oscillations $\omega _{\perp }^{2}\left(M,\rho \right)$ is given by Eq.~(\ref{Schfr}). By expressing the radial distance $\rho $ in terms of the gravitational radius $M$ so that $\rho =nM$, $n={\rm constant}\geq 6$, we obtain
\begin{equation}
\omega _{\perp }^2=\frac{1}{M^2\left[\left(n^2-2\right)\sqrt{1+n^2}-2\right]}.
\end{equation}

By introducing a set of dimensionless variables $\left(\theta ,\delta Z\right)$, defined as $t=\left(M/c\right)\theta $ and $\delta z=M\delta Z$, respectively, and by denoting the product $\nu M$ as $\zeta$, the equation of motion of the stochastic oscillating disk becomes
\begin{equation}
\frac{d^2\delta Z}{d\theta ^2}+\zeta \frac{d\delta Z}{d\theta }+\frac{1}{\left(n^2-2\right)\sqrt{1+n^2}-2}\delta Z=\eta ^{z}\left(\theta \right),
\end{equation}
where $\eta ^z\left(\theta \right)=M\xi^z (t)$.

The dimensionless luminosity $L_{\theta }$ can be written as
\begin{equation}
L_{\theta }=\zeta \left(\frac{d\delta Z}{d\theta }\right)^2-\frac{d\delta Z}{d\theta }\eta ^{z}\left(\theta \right),
\end{equation}
where $L_{\theta }=ML/c^3$.

\begin{figure}
 \centering
   \includegraphics[width=8cm]{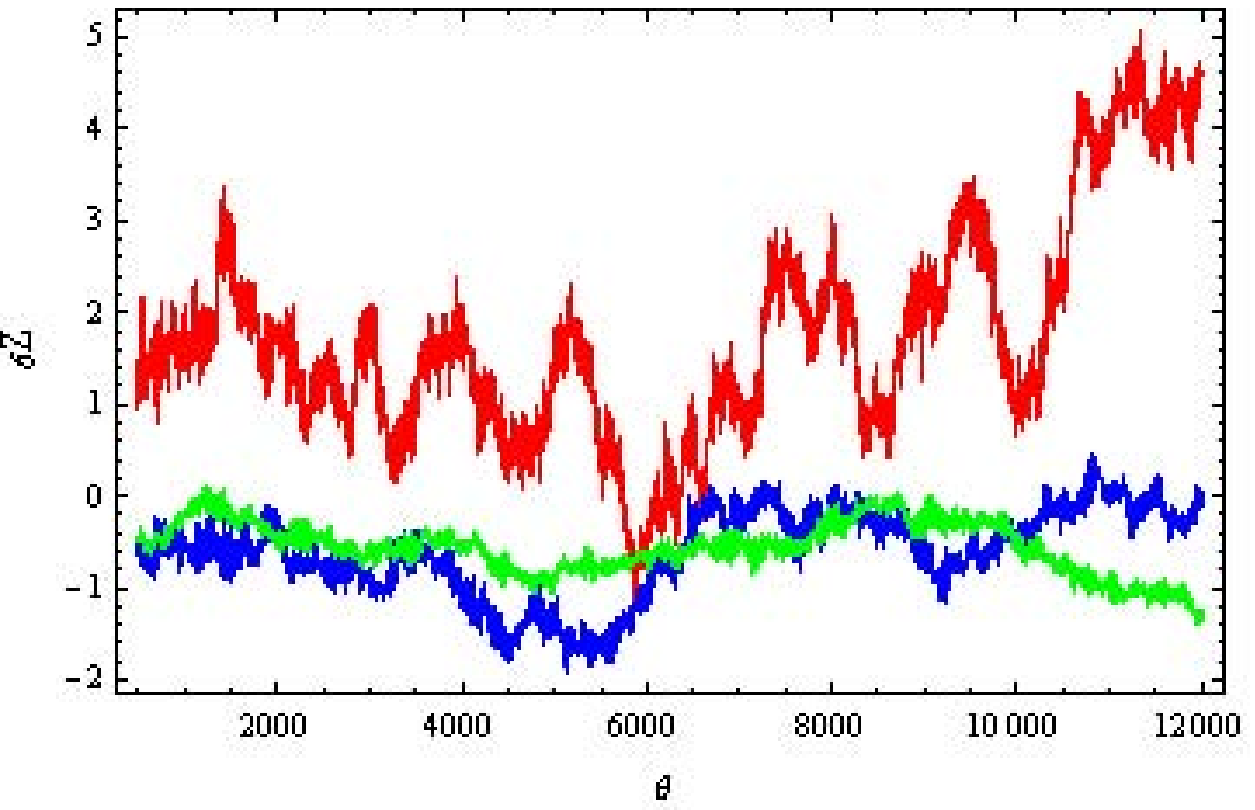}
   \includegraphics[width=8cm]{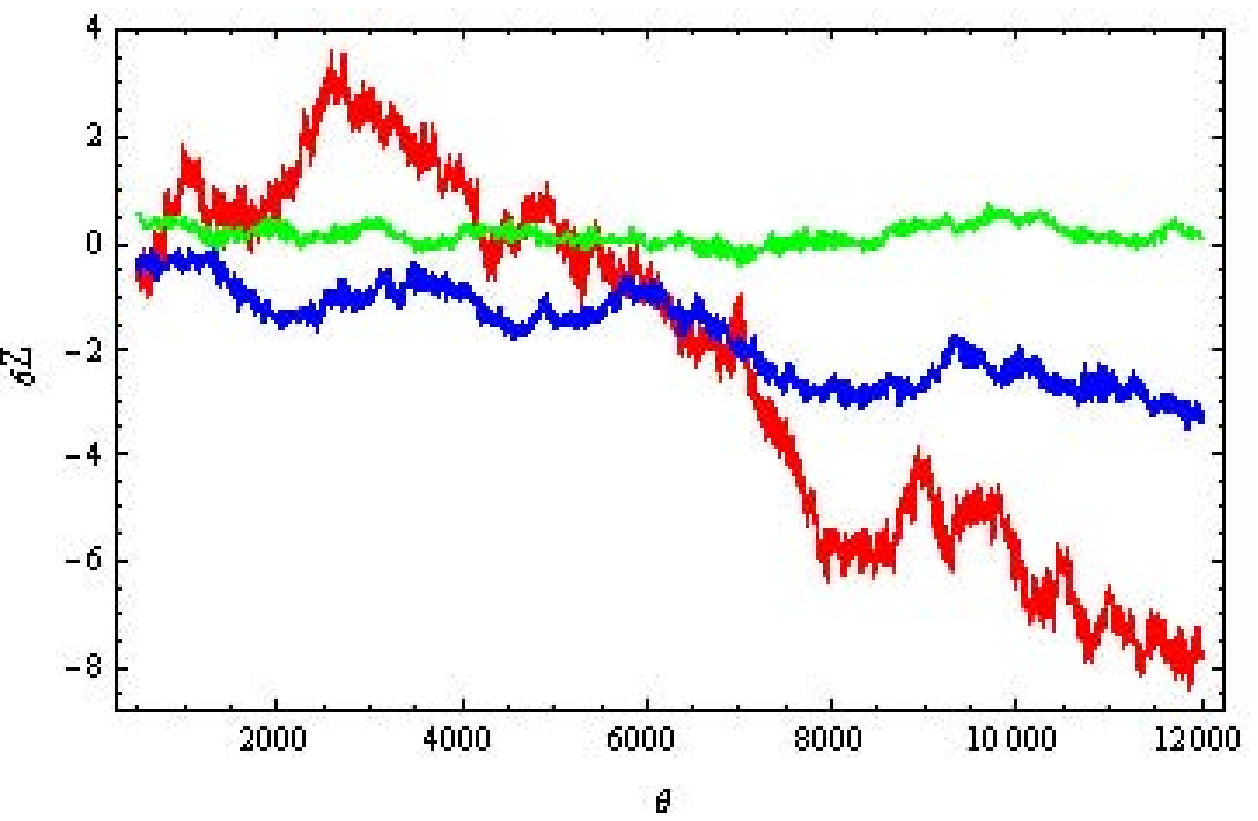}
      \caption{Variation of the dimensionless coordinate $\delta Z$ of the stochastically oscillating Schwarzschild disk as a function of $\theta $ for a central object with mass $M = M^{10}M_\odot$ and $\zeta = 100 $ (red curve), $\zeta = 2.5\times 100$ (blue curve), and $\zeta = 5\times 100$ (green curve).  In the upper Figure $n=7$, while in the lower Figure  $n \to \infty$.}
         \label{ZSch}
   \end{figure}

\begin{figure}
   \centering
   \includegraphics[width=8cm]{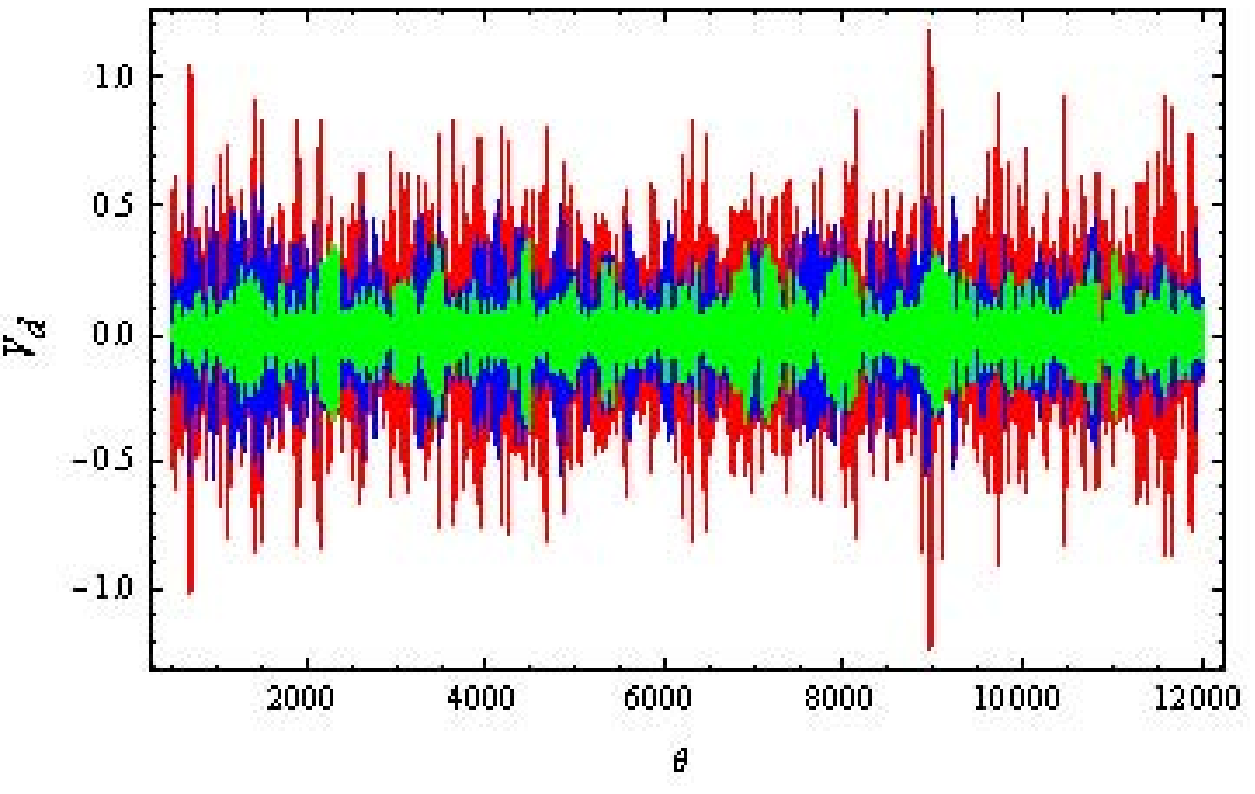}
    \includegraphics[width=8cm]{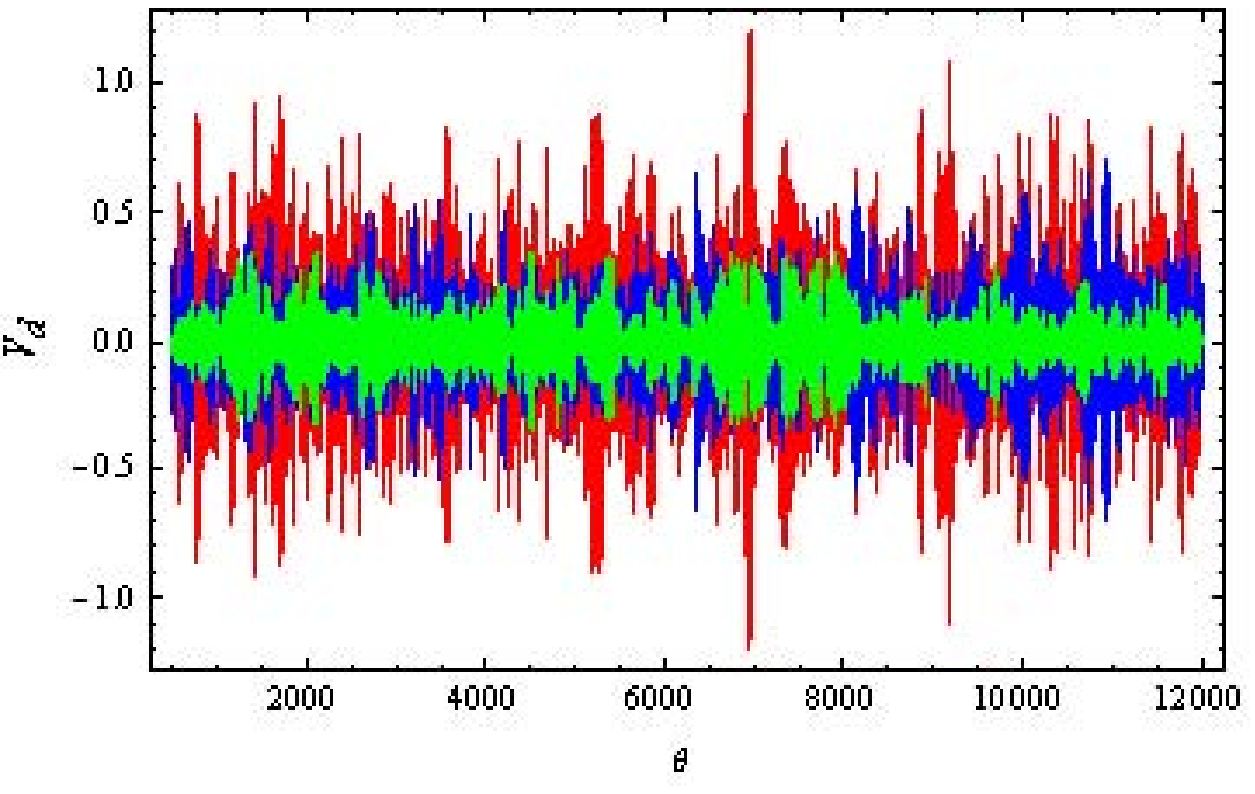}
      \caption{Variation of the dimensionless velocity $V_{d}=d\delta Z/\delta \theta $ of the stochastically oscillating Schwarzschild disk as a function of $\theta $ for a central object with mass $M = M^{10}M_\odot$ and for  $\zeta = 100 $ (red curve),  $\zeta = 2.5\times 100$ (blue curve), and $\zeta = 5\times 100$ (green curve), respectively. In the upper Figure $n=7$, while in the lower Figure  $n \to \infty$.}
         \label{VSch}
   \end{figure}

\begin{figure}
   \centering
   \includegraphics[width=8cm]{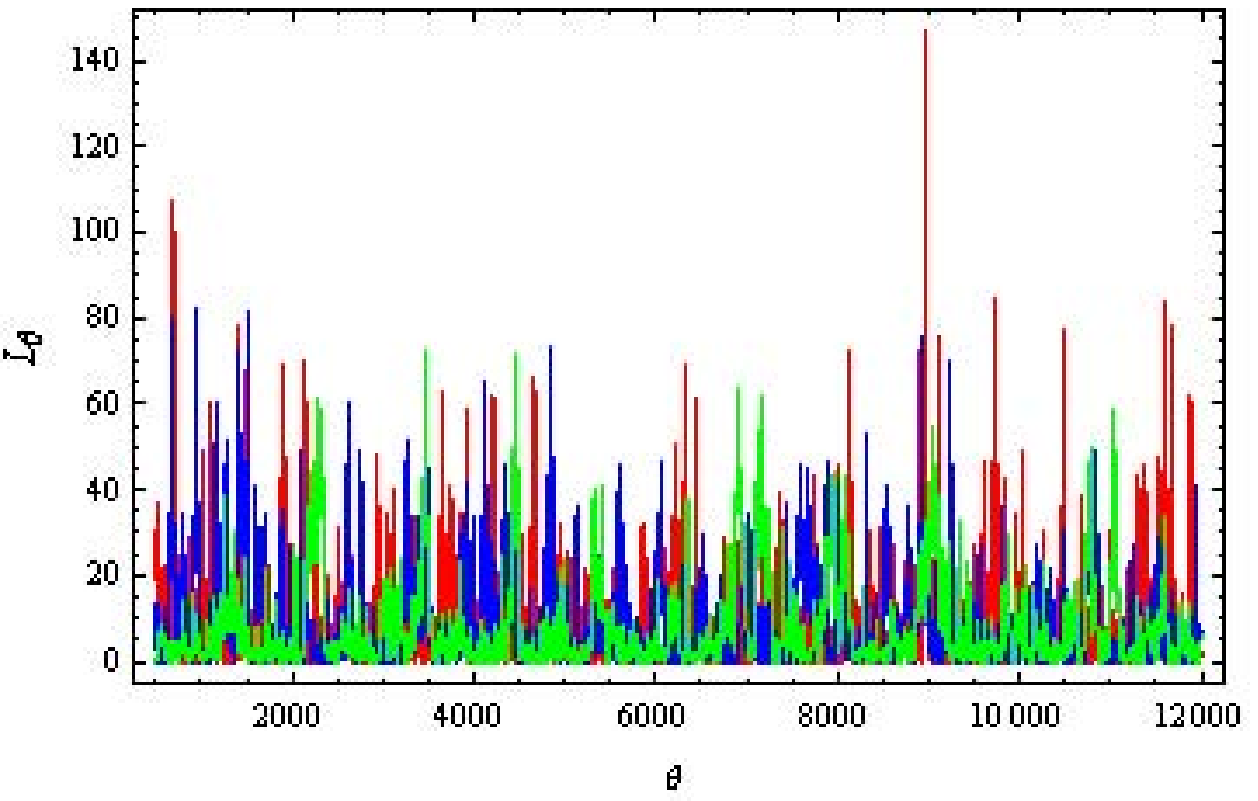}
    \includegraphics[width=8cm]{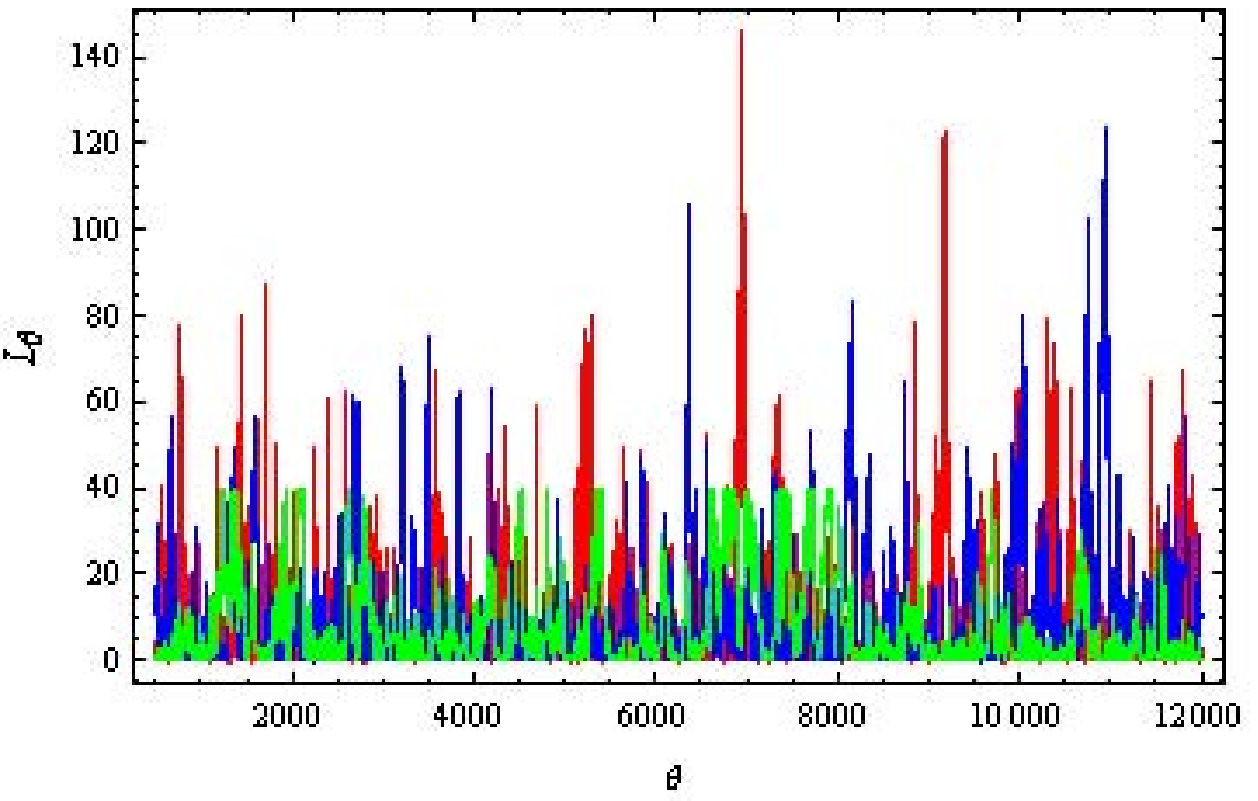}
      \caption{Dimensionless luminosity $L_{\theta }$ of the stochastically oscillating Schwarzschild disk for a central object with mass $M = M^{10}M_\odot$ and  $\zeta = 100 $ (red curve), $\zeta = 2.5\times 100$ (blue curve), and $\zeta = 5\times 100$ (green curve), respectively. In the upper Figure $n=7$, while in the lower Figure  $n \to \infty$.}
         \label{LSch}
   \end{figure}

The variations of the dimensionless coordinate $\delta Z$, of the stochastic velocity of the disk, and the luminosity of the stochastically perturbed disk are represented, for a central object with mass $M=10^{10}M_{\odot}$, for $n=7$ and $n\rightarrow\infty$, and for several values of $\zeta $, in Figs.~\ref{ZSch} - \ref{LSch}, respectively.

\begin{figure}
   \centering
   \includegraphics[width=8cm]{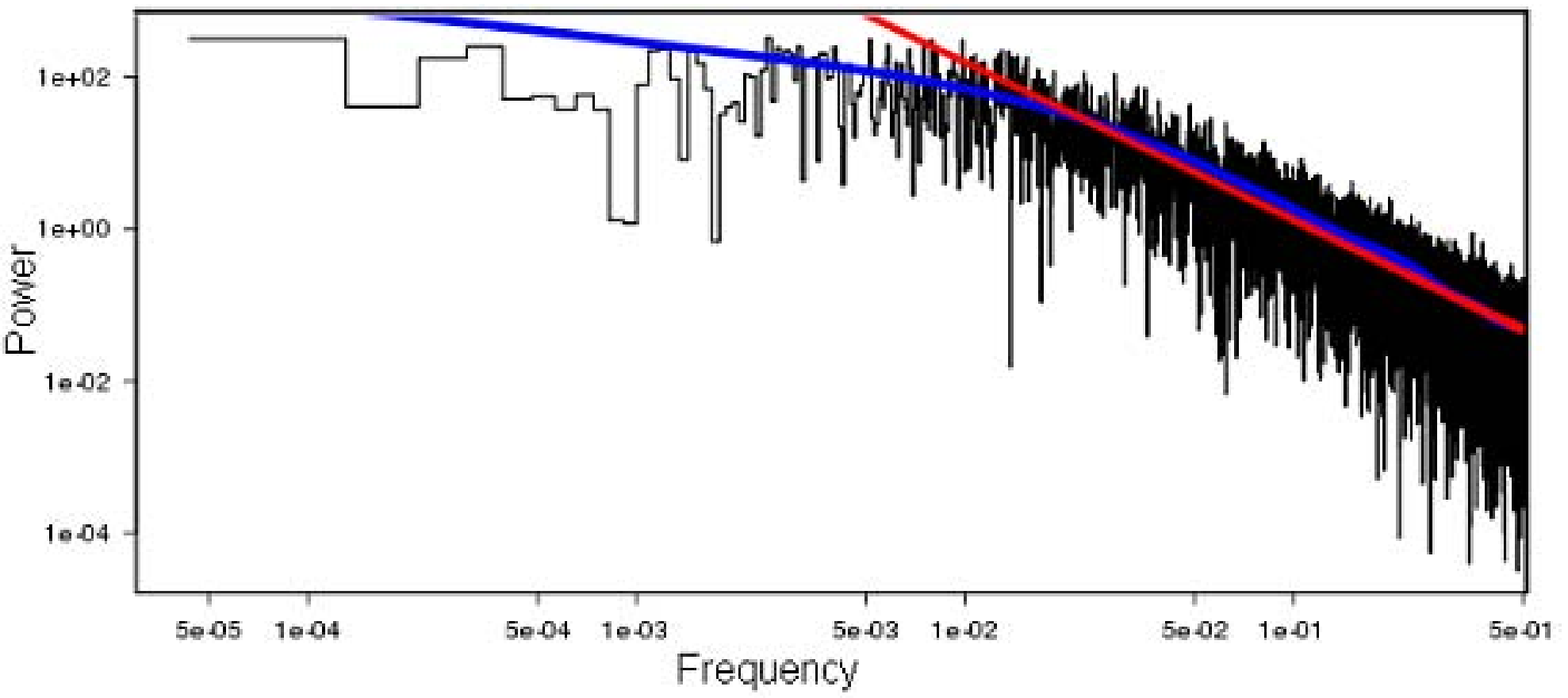}
    \includegraphics[width=8cm]{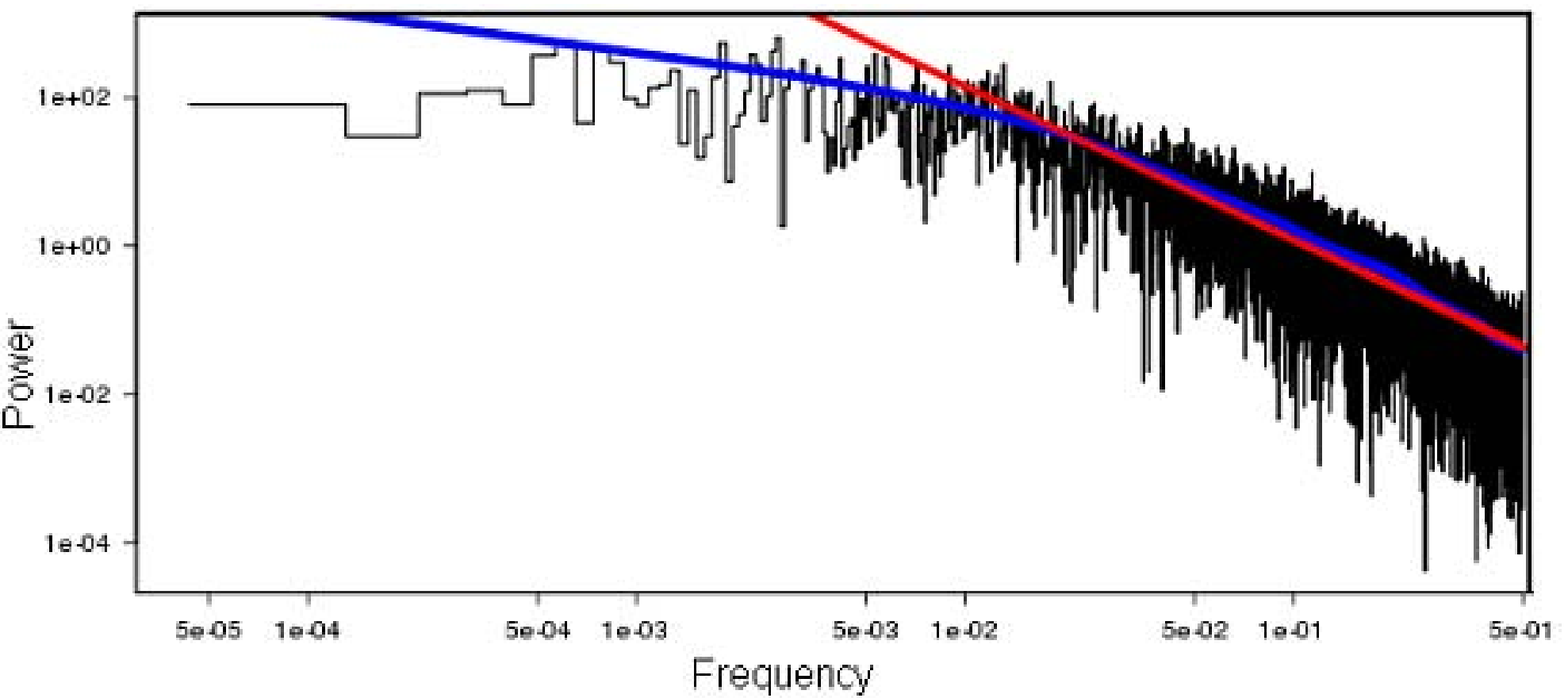}
      \caption{PSD of the luminosity of the stochastically oscillating Schwarzschild disk for a central object of mass  $M = M^{10}M_{\odot}$. In the upper Figure $n=7$ and $\alpha = 2.076$. In the lower Figure $n\rightarrow \infty$ and $\alpha = 2.070$. The red lines represent the fits according to $\mathcal{H}$.}
         \label{PSD1}
   \end{figure}

   In order to describe the stochastic characteristics of the physical parameters of the oscillating disk from a quantitative point of view we compute the Power Spectral Distribution (PSD) of the luminosity \citep{Vaughan} (for details see the Appendix). The numerical values of the slopes of the PSD give an insight into the nature of the mechanism leading to the observed variability. The computations of the PSD have been done by using the .R software and the assumption $\mathcal{H}:P(f)=\beta f^{-\alpha}$, where $\alpha $ and $\beta $ are constants, which was applied to the time series obtained by using the BBK integrator. The results for the Schwarzschild case are
shown in Fig.~\ref{PSD1}. In the case considered in the present paper, the luminosity is determined by the evolution, according to a Brownian-type equation of motion, of a volume element of the
matter in the disk. For such a situation we expect a PSD slope of $-2$, and this result in
fact represents a consistency check of the stochastic simulations.

  \subsection{The Kerr stochastic disk}

By denoting  $\rho = nM$, $a = k M$ and $\delta = M \sqrt{n^2+1-k^2}$, with $n>0$ and $k \in [0,1]$, the oscillation frequency of the Kerr disk, given by Eq.~(\ref{Kerrfr}), can be represented as
\begin{equation}
\omega _{\perp}^2=\frac{\omega _1^2\left(n,k\right)}{M^2\omega _2^2\left(n,k\right)},
\end{equation}
where
\begin{eqnarray}
\omega _{1}^{2}\left( n,k\right)  &=&3n^{4}+n^{2}\sqrt{1+\delta }\times
\nonumber \\
&&\left[ 2k\left( 5+3\delta \right) +3\sqrt{1+\delta }\left( 6+3\delta
+\delta ^{2}\right) \right] - \nonumber\\
&&2\left( 1+\delta \right) ^{3/2}\times \Big[ -4k(2+\delta )+\nonumber\\
&&\sqrt{1+\delta }%
\left( -8+3\delta ^{2}+\delta ^{3}\right) \Big] ,
\end{eqnarray}
and
\begin{eqnarray}
\omega _2^2\left(n,k\right) &=& \left(1+\delta \right) \left [ n^2 - \delta \left(1+\delta
\right) \left(3+\delta \right) \right ]^2 \times \Bigg \{
\delta-\nonumber\\
&&  \frac{\left [ k+ \left(1+\delta \right)^{3/2}\right
]^2 \left [ 4 \left(1+\delta \right) + n^2 \left(3+\delta
\right)\right]}{\left [ n^2 - \delta \left(1+\delta \right)
\left(3+\delta \right)\right]^2}+ \nonumber\\
&&\frac{4k\left[k +
\left(1+\delta \right)^{3/2}\right]}{n^2-\delta \left(1+\delta
\right) \left(3+\delta \right)}-1\Bigg\},
\end{eqnarray}
respectively.

Then the vertical dimensionless equation of motion of the stochastically oscillating Kerr disk is given by
\begin{equation}
\frac{d^2\delta Z}{d\theta ^2}+\zeta \frac{\delta Z}{d\theta }+\frac{\omega _1^2\left(n,k\right)}{\omega _2^2\left(n,k\right)}\delta Z=\eta ^{z}\left(\theta \right).
\end{equation}

\begin{figure}
   \centering
   \includegraphics[width=8cm]{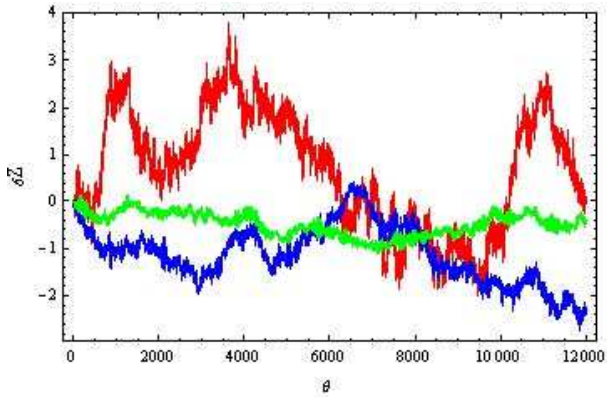}
      \caption{Variation of the dimensionless coordinate $\delta Z$ of the stochastically oscillating disk as a function of $\theta $ for a rotating Kerr black hole with mass $M = M^{10}M_\odot$ and $a=0.9M$, for $n=7$, and for different values of $\zeta $:  $\zeta = 100 $ (red curve),  $\zeta = 2.5\times 100$ (blue curve), and $\zeta = 5\times 100$ (green curve), respectively.}
         \label{ZKerr}
   \end{figure}

\begin{figure}
   \centering
   \includegraphics[width=8cm]{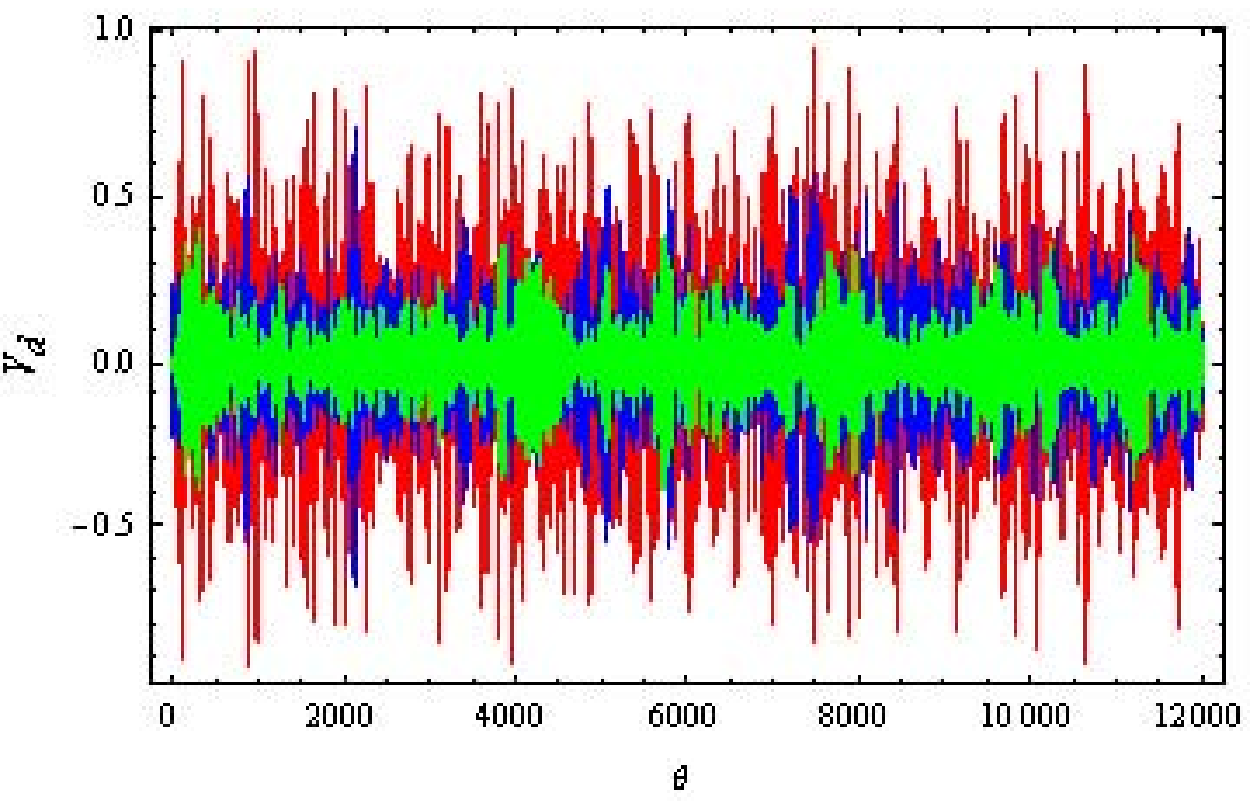}
   \caption{Variation of the dimensionless velocity $V_{d}=d\delta Z/d {\theta }$ of the stochastically oscillating disk as a function of $\theta $ for a rotating Kerr black hole with mass $M = M^{10}M_\odot$ and $a=0.9M$, for $n=7$, and for different values of $\zeta $:  $\zeta = 100 $ (red curve), $\zeta = 2.5\times 100$ (blue curve), and $\zeta = 5\times 100$ (green curve), respectively.}
         \label{VKerr}
   \end{figure}

\begin{figure}
   \centering
   \includegraphics[width=8cm]{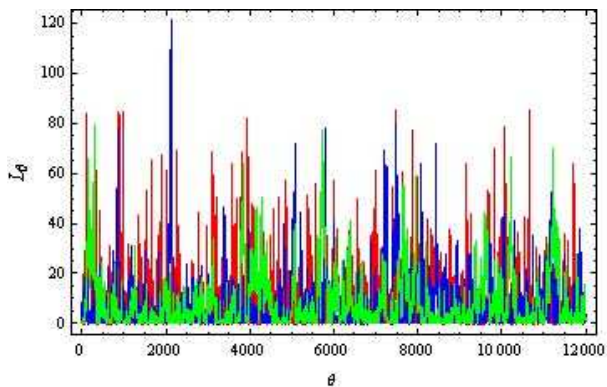}
   \caption{Variation of the luminosity $L_{\theta} $ of the stochastically oscillating disk as a function of $\theta $ for a rotating Kerr black hole with mass $M = M^{10}M_\odot$ and $a=0.9M$, for $n=7$, and for different values of $\zeta $:  $\zeta = 100 $ (red curve), $\zeta = 2.5\times 100$ (blue curve), and $\zeta = 5\times 100$ (green curve), respectively.}
         \label{LKerr}
   \end{figure}

In Figs.~\ref{ZKerr}-\ref{LKerr} we have represented the
variations of the dimensionless coordinate $\delta Z$, of the
velocity $V_{d}=d\delta Z/d \theta $, and of the luminosity for a
stochastically oscillating Kerr disk for a rotating massive
central object with mass $M=10^{10}M_{\odot}$ and $a=0.9M$, for
$n=7$, and for different values of $\zeta$.

\begin{figure}
   \centering
   \includegraphics[width=8cm]{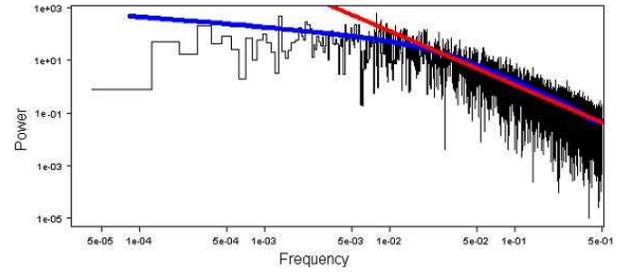}
      \caption{PSD of the luminosity for a stochastically oscillating disk around a Kerr black hole with mass $M = M^{10}M_\odot$ and $a=0.9M$, for $n=7$. The slope of the PSD is $2.046$.}
         \label{PSD2}
   \end{figure}

 In Fig.~\ref{PSD2} we show the PSD of the luminosity of the stochastically oscillating disk around a Kerr black hole with mass $M=10^{10}M_{\odot}$ and $a=0.9M$, respectively.

\section{Discussions and final remarks}\label{s8}

In the present paper we have introduced a mathematical model for the description of the perturbations of a thin accretion disk in contact with an exterior stochastic medium (a thermal bath). As a result of the interaction between the medium and the disk, the motion of the particles in the disk has essentially a stochastic nature. To obtain the equations of motion of the perturbed disk we have generalized the approach introduced in \citet{Shi} and \citet{Sem}, by introducing in the equations of the perturbed geodesic lines a dissipative viscous term, and a stochastic force term, respectively. The equations of the motion of the fluctuating disk have been obtained in both the equatorial and the vertical plane, as well as the general stochastic energy transfer equation. We have studied in detail the vertical oscillations of the disk, and we have shown that in this case the motion is described by a standard Langevin type differential equation for an harmonic oscillator, with a geometry dependent oscillation frequency. By numerically integrating the equation of motion we have obtained the vertical displacements, velocities and the luminosity of the disk in the cases of both Schwarzschild and Kerr geometries.

 Fluctuations of accretion disks around super-massive black holes may have important astrophysical applications. The emission of Galactic Black Hole Binaries (BHBs) and Active Galactic Nuclei (AGN) displays a significant aperiodic variability on a broad range of time - scales. The Power Spectral Density  of such variability is generally modeled
with a power law, $P(f)\propto f^{-\alpha }$, where  power-law index $\alpha $ keeps a constant value in a certain range of $f$, but changes among different ranges. At high frequencies, the PSDs of both BHBs and
AGNs present a steep slope with $\alpha \sim 2$. On the contrary, below a break frequency, typically at a few Hz for BHBs, they flatten to a slope with $\alpha \sim 1$, representing the so-called flicker noise \citep{King}. However, the observed power spectra often deviate from
the form $f^{-1}$. For example, the PSD of Cyg X-1 is well described with the form $f^{-1}$ in the soft state, but in the hard state it exhibits the form $f^{-1.3}$ \citep{Gilf}. In particular, it was shown that the power-law index is around $\alpha =0.8 - 1.3$ both in the soft state of BHBs and in narrow-line Seyfert 1 galaxies \citep{Janiuk}.

In the case of the present model of the disk oscillations under the effect of some stochastic forces $\xi (t)$, the resulting Power Spectral Density is of the form $P(f)\propto f^{-2}$. This result can be understood qualitatively by noting that $P(f)\propto \int{\exp\left(ift\right)\left[\int^t{\xi \left(t'\right)dt'}\right]dt}\propto if^{-1}\int ^t{\exp\left(ift\right)\xi (t)dt}$, which gives $P\propto f^{-2}$, since $\xi (t)$ is a stochastic variable. This result is generally also valid for harmonically oscillating, undamped disks, and consequently it is independent of the general relativistic corrections to the equation of motion, since these only change the oscillation frequency of the compact object - disk system. However, as pointed out previously,  observations of BHBs and AGNs have shown that the spectral index is not $\alpha =2$, but it is either $\alpha \approx 1$, or, more generally, $\alpha \in \left(0.8,2\right)$. Such a dispersion of the power-law index reveals that there must exist some other mechanisms, different from purely stochastic oscillations, which  may be responsible for the observed value of $\alpha $, like, for example,  hydrodynamic fluctuations, magnetohydrodynamic turbulence, magnetic flares, density fluctuations in the corona, or variations of the accretion rate, caused by small amplitude variations in the viscosity \citep{King}.

In a model introduced in \citet{Min2} aperiodic X-ray fluctuations are thought to originate from instabilities of the accretion disk around a black hole. To describe these type of fluctuations a cellular automaton model was proposed  \citep{Min1}. In this model, a gas particle is randomly injected into an accretion disk surrounding a black hole. When the mass density of the disk exceeds some critical value at a certain point, an instability develops, and the accumulated matter begins to drift inward as an avalanche, thereby emitting X-rays.  Within the framework of this  model, one can generate $ f^{-1}$ - like fluctuations in the X-ray luminosity, in spite of the random mass injection. A similar model can be used to explain the optical-ultraviolet "flickering" variability in cataclysmic variables \citep{Min3}. Light fluctuations are produced by occasional flare-like events, and subsequent avalanche flow in the accretion disk atmospheres.

Stochastic oscillations of the accretion disks can also provide an alternative model for the explanation of the observed intra-day variability in BL Lac objects, and for other similar transient events \citep{Leung}. The basic physical idea of this model is that the source of the intra-day variability can be related to some stochastic oscillations of the disk, triggered by the interaction of the disk with the central supermassive black hole, as well as with a background cosmic environment, which perturbs the disk.  To explain the observed light curve behavior,  a model for the stochastic oscillations of the disk was developed, by taking into account the gravitational interaction with the central object, the viscous type damping forces generated in the disk, and a stochastic component which describes the interaction with the cosmic environment. The stochastic oscillation model can reproduce the aperiodic light curves associated with transient astronomical phenomena.

Random or fluctuating phenomena can be found in many  natural processes. In the present paper we have investigated the stochastic  properties of the oscillating general relativistic thin accretion disks, and we have introduced an  approach in which the stochastic processes related to disk instabilities can be described in a unitary approach by a Langevin type equation that includes both the deterministic, general relativistic,  and the random forces acting on the accretion disks. Our results can be considered only as a first step in the investigation of the stochastic properties of the general relativistic accretion disks. In order to obtain a more accurate determination of the relevant physical parameters of the disks, and to determine more exactly the possible astrophysical applications of the model, it would be important to also investigate the radial stochastic oscillations of the disks around compact general relativistic objects.

\section*{Acknowledgments}

We would like to thank to the anonymous referee for comments and suggestions that helped us to significantly improve the manuscript. The work by T. H. was supported by a grant from the Research Grants Council of the Hong Kong Special
Administrative Region, China. G. M. acknowledges the financial support of the Sectoral Operational Programme for Human Resources Development 2007-2013, co-financed by the European Social Fund, under the project number POSDRU/107/1.5/S/76841 with the title "Modern Doctoral Studies: Internationalization and
Interdisciplinarity". G. M. would like to thank Vienna University of Technology, Institute of Theoretical Physics, for its hospitality during the time when most of this work was written.

\section*{Appendix}

In statistical analysis, if $X$ is some fluctuating quantity, with
mean $\mu$ and variance $\sigma ^2$, then a correlation function
for quantity $X$ is defined as

\[
R(\tau) = \frac{\langle \left ( X_s - \mu \right ) \left (
X_{s+\tau} - \mu \right) \rangle}{\sigma ^2}.
\]
where $X_s$ is the value of $X$ measured at time $s$ and $\langle
\rangle$ denotes averaging over all values $s$. Based on the
correlation function the Power Spectral Distribution (PSD) is
defined  as
\[
P(f) = \int _{-\infty} ^{+\infty} R(\tau) e^{-\imath 2\pi f \tau} .
\]

It is straightforward to see the importance of the PSD in terms of the
"memory" of a given process. For example, if $X$ is the measured
luminosity of a perturbed disk, the slope of the PSD of a time
series of $X$ provides an insight into the degree of correlation the
underlying physical processes have with themselves. The system needs
additional energy to fluctuate, and this mechanism is
best explained for the Brownian motion, in which case the energy is
thermal. Brownian motion produces a PSD of the form $P(f) \sim f^{-2}$.

However, assessment of the PSD slope from one single observational
time-series is non-trivial. In such cases it is very helpful to
use dedicated software, such as the statistical software .R
\citep{Vaughan}. Basically, the input to this software consists
of one time series. Any assumption about the behaviour of the
source that produced this time series may be expressed in the
form of an analytical function. For example, we want to test the
assumption that the source producing the time series behaves so as
to be described by
\[
\mathcal{H}:P(f)=\beta f^{-\alpha}.
\]

The software will return the most probable values of the parameters
$\{\alpha, \beta \}$, and the probability that the source indeed
behaves according to this hypothesis.


\begin{thebibliography}{99}

\bibitem[\protect\citeauthoryear{Binni et al.}{2005}]{Bi05}  Bini D., De Paolis F., Geralico A.,
Ingrosso G. \&  Nucita A., 2005, Gen. Rel. Grav., 37, 1263

\bibitem[\protect\citeauthoryear{Blaes et al.}{2006}]{Bla}  Blaes O. M., Arras P., \& Fragile P. C., 2006, MNRAS, 369,  1235

\bibitem[\protect\citeauthoryear{Br\"unger et al.}{1984}]{bbk} Br\"unger A., Brooks C. L., \& Karplus M., 1984, Chem. Phys. Lett., 105, 495

\bibitem[\protect\citeauthoryear{Chatterjee et al.}{2002}]{Chat} Chatterjee P., Hernquist L., \&  Loeb A., 2002,  Astrophys. J., 572,  371

\bibitem[\protect\citeauthoryear{Chavanis \& Sire}{2004}]{Cha} Chavanis P.-H. \&  Sire, C., 2004, Phys. Rev. E, 69, 016116

\bibitem[\protect\citeauthoryear{Coffey et al.}{2004}]{Co04} Coffey W. T.,  Kalmykov Yu. P., \& Waldron J. T., 2004,  The Langevin Equation, with Applications to Stochastic Processes in Physics,
Chemistry and Electrical Engineering, World Scientific, New Jersey

\bibitem[\protect\citeauthoryear{Dunkel \& H\"anggi }{2005a}]{Du05a} Dunkel J. \& H\"anggi P., 2005, Phys. Rev. E, 71, 016124

\bibitem[\protect\citeauthoryear{Dunkel \& H\"anggi }{2005b}]{Du05b} Dunkel J. \& H\"anggi P., 2005, Phys. Rev. E, 72, 036106

\bibitem[\protect\citeauthoryear{Dunkel \& H\"anggi }{2009}]{Du09} Dunkel J. \& H\"anggi P., 2009, Phys. Rep. 471(1), 1

\bibitem[\protect\citeauthoryear{Fan et al. }{2008}]{Fan}
    Fan J. H., Rieger F. M., Hua T. X., Joshi, U. C., Li J., Wang Y. X., Zhou, J. L., Yuan Y. H., Su, J. B., \& Zhang Y. W., 2008, Astropart. Phys.,  28,  508

\bibitem[\protect\citeauthoryear{Gilfanov }{2010}]{Gilf} Gilfanov M., 2010, The Jet Paradigm, Lecture Notes in Physics, Volume 794, Springer-Verlag Berlin, Heidelberg, p. 17

\bibitem[\protect\citeauthoryear{Hawking \& Ellis }{1973}]{Hawk} Hawking S. W. \& Ellis G. F. R., 1973, The large scale structure of space-time, London, University Press

\bibitem[\protect\citeauthoryear{Horak et al. }{2009}]{Hor} Horak J., Abramowicz M. A., Kluzniak W., Rebusco P., \& Torok G., 2009, Astron. Astrophys., 499, 535

\bibitem[\protect\citeauthoryear{Hu \& Verdaguer }{2008}]{stoch}  Hu B. L. \& Verdaguer E., 2008, Living Rev. Relativity, 11, 3

\bibitem[\protect\citeauthoryear{Janiuk \& Czerny}{2007}]{Janiuk}  Janiuk A. \& Czerny B., 2007, Astron. Astrophys., 466, 793

\bibitem[\protect\citeauthoryear{Kato}{2001}]{Kato1} Kato S., 2001, PASJ, 53, 1

\bibitem[\protect\citeauthoryear{King et al.}{2004}]{King} King A. R., Pringle J. E., West R. G., \& Livio M., 2004,
MNRAS, 348, 111

\bibitem[\protect\citeauthoryear{Landau \& Lifshitz}{1987}]{Lan} Landau L. D. \& Lifshitz E. M., 1987, Fluid mechanics, Oxford, England, Pergamon Press

\bibitem[\protect\citeauthoryear{Lazarian \& Roberge}{1997}]{Laz}  Lazarian A. \& Roberge W. G., 1997, Astrophys. J., 484, 230

\bibitem[\protect\citeauthoryear{Leung et al.}{2011}]{Leung}  Leung C. S., Wei J. Y., Harko T., \& Kovacs Z., 2011, J. Astrophys. Astron., 32,  189

\bibitem[\protect\citeauthoryear{Mineshige et al.}{1994}]{Min1} Mineshige S., Ouchi N. B., \& Nishimori H., 1994, PASJ, 46, 97

    \bibitem[\protect\citeauthoryear{Okazaki et al.}{1987}]{Okaz}  Okazaki, A. T., Kato, S., \& Fukue, J., 1987, PASJ, 39, 457

\bibitem[\protect\citeauthoryear{O'Neill et al.}{2009}]{On} O'Neill S. M., Reynolds C. S., \& Miller M. C., 2009,  Astrophys. J., 693,  1100

\bibitem[\protect\citeauthoryear{Perez et al.}{1997}]{Per}  Perez C. A., Silbergleit A. S., Wagoner R. V., \& Lehr D. E., 1997, Astrophys. J., 476, 589


\bibitem[\protect\citeauthoryear{Pope}{2007}]{Pope} Pope, E. C. D., 2007, MNRAS, 381,  741

\bibitem[\protect\citeauthoryear{Rezzolla et al.}{2003}]{Rez}   Rezzolla L., Yoshida S., \& Zanotti O., 2003, MNRAS, 344,  978

\bibitem[\protect\citeauthoryear{Rodriguez et al.}{2002}]{Rod} Rodriguez M. O., Silbergleit A. S., \& Wagoner R. V., 2002, Astrophys. J., 567, 1043


\bibitem[\protect\citeauthoryear{Semerak \& Zacek}{2000}]{Sem} Semerak O. \& Zacek M., 2000, Publ. Astron. Soc. Japan, 52, 1067

  \bibitem[\protect\citeauthoryear{Shi \& Li}{2000}]{Shi1}  Shi C.-S. \& Li X.-D., 2010,  Astrophys. J.,  714,  1227

\bibitem[\protect\citeauthoryear{Shirokov}{1973}]{Shi} Shirokov M. F., 1973, Gen. Rel. Grav., 4, 131

\bibitem[\protect\citeauthoryear{Silbergleit et al.}{2001}]{Sil} Silbergleit, A. S., Wagoner, R. V., \& Rodriguez, M., 2001, Astrophys. J., 548, 335

\bibitem[\protect\citeauthoryear{Skeel \& Izaguirre}{2002}]{Skeel} Skeel R. D. \& Izaguirre J. A., 2002, Molecular Physics, 100, 3885

\bibitem[\protect\citeauthoryear{Tagger \& Varniere}{2006}]{Tag} Tagger M. \& Varniere P., 2006, Astrophys. J.,  652,  1457

\bibitem[\protect\citeauthoryear{Takeuchi \& Mineshige}{1997}]{Min2} Takeuchi M. \& Mineshige S., 1997, Astrophys. J., 486, 160

\bibitem[\protect\citeauthoryear{Titarchuk \& Osherovich}{2000}]{Tit} Titarchuk L. \& Osherovich V., 2000, Astrophys. J., 542,  L111

\bibitem[\protect\citeauthoryear{Vietri et al.}{2003}]{Vietri} Vietri M.,  De Marco D., \& Guetta D., 2003, Astrophys. J., 592, 378

\bibitem[\protect\citeauthoryear{Vaughan}{2010}]{Vaughan} Vaughan, S., 2010, MNRAS, 402,
307

\bibitem[\protect\citeauthoryear{Weinberg}{1972}]{Wein} Weinberg S., 1972, Gravitation and cosmology: principles and applications of the general theory of relativity, New York, Wiley

 \bibitem[\protect\citeauthoryear{Yonehara et al.}{1997}]{Min3} Yonehara A., Mineshige S., \& Welsh W. F., 1997,
 Astrophys. J., 486, 388

\bibitem[\protect\citeauthoryear{Zanotti et al.}{2005}]{Zan} Zanotti O, Font J. A., Rezzolla L., \& Montero P. J., 2005, MNRAS, 356,  1371
\end{thebibliography}
\end{document}